\documentclass[preprint,aps,prd,amsmath,amssymb,nofootinbib]{revtex4-1}
\usepackage{graphicx}
\usepackage{color}
\usepackage[latin1]{inputenc}
\usepackage{ulem}
\newcommand{\be}{\begin{eqnarray}}
\newcommand{\beq}{\begin{equation}}
\newcommand{\eeq}{\end{equation}}
\newcommand{\ee}{\end{eqnarray}}

\newcommand{\bmp}{\noindent\begin{minipage}{16cm}}
\newcommand{\emp}{\end{minipage}\vskip 7mm} 

\usepackage{graphicx}
\usepackage{dcolumn}
\usepackage{bm}
\usepackage{amsmath}
\usepackage{amsfonts}

\def\drawbox#1#2{\hrule height#2pt
        \hbox{\vrule width#2pt height#1pt \kern#1pt
              \vrule width#2pt}
              \hrule height#2pt}

\def\Asym#1#2{\vcenter{\vbox{\drawbox{#1}{#2}
              \kern-#2pt 
              \drawbox{#1}{#2}}}}

\def\({\left(}
\def\){\right)}

\begin{document}

\title{Constraining Asymmetric Dark Matter through observations of compact stars}
\author{Chris {\sc Kouvaris}}\email{kouvaris@cp3.sdu.dk}
\affiliation{$\text{CP}^3$-Origins, University of Southern Denmark, Campusvej 55, Odense 5230, Denmark}
\author{Peter {\sc Tinyakov}}\email{Petr.Tiniakov@ulb.ac.be}
 \affiliation{Service de Physique Th\'eorique,  Universit\'e Libre de Bruxelles, 1050 Brussels, Belgium}
 \preprint{CP3-Origins-2010-54}
 
\begin{abstract}
We put constraints on asymmetric dark matter candidates with spin-dependent  interactions based on the simple existence of 
white dwarfs and neutron stars in globular clusters. For a wide range of the parameters (WIMP mass and WIMP-nucleon cross section),
 WIMPs can be trapped in progenitors in large numbers and once the original star collapses to a white dwarf or a neutron star, these WIMPs might self-gravitate
 and eventually collapse forming a mini-black hole that eventually destroys the star. We 
 impose constraints competitive to direct dark matter search experiments, for WIMPs with masses down to the TeV scale.

\end{abstract}


\maketitle
\section{Introduction}

Observations of clusters of galaxies, rotations curves of individual
galaxies, cosmic microwave background anisotropies, and many other
methods suggest the existence of dark matter. A possible realization
of dark matter might be in the form of Weakly Interacting Massive
Particles (WIMPs). A huge effort is being undertaken by
experimentalists to directly detect WIMPs in underground or space
experiments, as well as by theorists to incorporate them into viable
theories beyond the Standard Model.

The situation experimentally is still not clear, as the majority of
the experiments have not detected WIMPs so far. Direct search
experiments with Earth based detectors like
CDMS~\cite{Ahmed:2009zw} and Xenon~\cite{Angle:2008we} have imposed
constraints on the WIMP-nuclei cross sections, assuming the local dark
matter density around the Earth as inferred from the cosmological and
other data (see e.g. Ref.~\cite{Dunkley:2008ie} for the
determination of the amount of dark matter from the WMAP data). On the
other hand, DAMA experiment~\cite{Bernabei:2010mq} claims dark
matter detection with parameters that contradict other experiments
if taken at face value.

Given the still unclear picture regarding the nature of dark matter,
it is of crucial importance to constrain as much as possible the WIMP
candidates, including their mass and interactions. Several such 
 candidates exist in the market depending on what theory beyond
the Standard Model one chooses, ranging from supersymmetry
~\cite{Jungman:1995df,Bertone:2004pz} and  hidden
sectors~\cite{Pospelov:2007mp,Hambye:2008bq}, to 
Technicolor~\cite{Gudnason:2006yj,Kouvaris:2007iq,Belotsky:2008vh,Kouvaris:2008hc,Ryttov:2008xe,Frandsen:2009mi,Kainulainen:2010pk}.
The WIMPs can be classified according to their properties, i.e. if
they are produced thermally, if they are
asymmetric~\cite{Barr:1990ca,Gudnason:2006yj,Belyaev:2010kp}, if they
have spin-dependent or spin-independent cross section with the
nuclei, if their collisions with the
nuclei are elastic or inelastic~\cite{TuckerSmith:2001hy,TuckerSmith:2004jv,Fargion:2005ep,Khlopov:2007ic,Khlopov:2008ty},
and/or whether they are
self-interacting~\cite{Spergel:1999mh,Dave:2000ar,Zentner:2009is,Frandsen:2010yj}. 

Apart from direct searches, constraints on the properties of the WIMPs
might arise from astrophysical observations as for
example in~\cite{Frandsen:2010mr}.  Concentration of the WIMPs within
stars can affect, under certain circumstances, the evolution of the
latter, and/or products of WIMP annihilation within the stars could be
directly or indirectly detected. The capture of WIMPs in the Sun and
the Earth~\cite{Press:1985ug,Gould:1987ju,Gould:1987ww} has been used
to predict a possible signature for an indirect detection of dark
matter based on neutrino production due to WIMP
co-annihilation~\cite{Jungman:1994jr,Nussinov:2009ft}. Constraints on
the dark matter properties and the dark matter profile can also be
imposed due to the effect of dark matter on the evolution of low mass
stars~\cite{Casanellas:2009dp,Casanellas:2010sj}, and main sequence stars~\cite{Scott:2008ns},
on possible
gravitational collapse of neutron stars~\cite{Goldman:1989nd}, and on
the cooling process of compact objects such as neutron stars and white
dwarfs
~\cite{Kouvaris:2007ay,Sandin:2008db,Bertone:2007ae,McCullough:2010ai,Kouvaris:2010vv,deLavallaz:2010wp}.
In particular, the authors of~\cite{Goldman:1989nd} have investigated
under what conditions a neutron star can collapse gravitationally due
to accretion of WIMPs, providing an upper bound for the WIMP masses
given the local dark matter density and the time of
accretion. Although the bound for the mass of a bosonic WIMP was low
$\sim 10$~MeV, the upper bound for fermionic WIMPs was quite high
$\sim 10^5$~TeV (and therefore not relevant for physics at the TeV
scale).

In this paper we investigate possible constraints that can arise from
stars that accrete WIMPs during their lifetime and then collapse into
a more compact object, white dwarf or a neutron star, inheriting the
accumulated dark matter. Depending on the location of the star and the
WIMP-nuclei cross section, it might be possible to impose constraints
on the mass of the WIMP, excluding in some cases candidates that are
lighter than TeV. Such constraints improve significantly the
existing ones, and may become relevant for  LHC physics.

More specifically, we consider two different cases. In the first
case, we examine the accretion of WIMPs with spin-dependent
interactions with protons onto a Sun-like star. We deduce under what
circumstances the accumulated WIMPs can trigger a gravitational
collapse once the star has turned into a white dwarf. In the second
 case, a supermassive star accretes
WIMPs which have a spin-dependent cross section with nucleons (protons
and neutrons).  A typical supermassive star of 15 solar masses lives
about $10^7$~yr and then explodes forming a neutron star. Under certain assumptions, the WIMPs
inherited by the neutron star from its progenitor will thermalize,
sink to the center and, for some range of parameters, collapse further
into a black hole. Thus, the mere existence of neutron stars might
impose constraints on the mass or the cross section of the WIMPs.  In
all cases we will assume cross sections that are compatible with
experimental constraints from the direct dark matter searches.

\section{Accretion of WIMPs onto Stars}

The crucial parameter that determines the capture of WIMPs by a star
is the WIMP-nucleon cross section. In this paper we will consider
dark matter candidates that exhibit predominantly a spin-dependent  cross section with nucleons. Spin-independent cross
section is highly constrained from direct dark matter search
experiments, whereas the spin-dependent one is less constrained. The
present constraint on the spin-independent cross section (normalized
to a single proton) is roughly $3\times 10^{-44}\text{cm}^2$
for a WIMP of a mass 100 GeV~\cite{Ahmed:2009zw}. However, the rate of
events is proportional to the local WIMP number density, and because
for a fixed dark matter energy density the number density decreases
inversely proportional to the WIMP mass, the present constraint can be
written (for masses higher than TeV) as
\[
\sigma_{\rm SI}  < 3 \times 10^{-43}\text{cm}^2 \left( {m\over
\text{TeV}}\right).
\]
As one can see, the constraint becomes weaker at higher masses. 

The constraints on spin-dependent cross section are not so strict. The
best upper limit on the spin-dependent cross section of WIMP with
neutron is $\sim 10^{-38}\text{cm}^2$ ~\cite{Ahmed:2008eu}, whereas
the analysis for spin-dependent cross section between WIMP and proton
gives a minimum upper limit by one order of magnitude
higher~\cite{Kopp:2009qt}. The constraint for a spin-dependent cross
section between WIMP-nucleon (proton or neutron) can be written as
\[
\sigma_{\rm SD} < 7\cdot 10^{-38}\text{cm}^2 
\left(m\over\text{TeV}\right).
\]
This constraint also weakens linearly in $m$ at high masses.  

The difference between the spin-independent and spin-dependent
constraints is due to several reasons. The first one is that WIMPs
with the spin-independent cross section scatter coherently with the
whole nucleus if their De~Broglie wavelength is larger than the size
of the nucleus. This condition is easily met on Earth based
detectors. The coherence increases the cross section between WIMP and
nucleus compared to the one of WIMP-nucleon roughly by a factor of
$N^2$ where $N$ is the number of nucleons composing the nucleus. For
example, this gives an enhancement by a factor of $\sim 73^2$ in the
case of the Ge detectors of CDMS. In addition, form factors suppress
further the spin-dependent cross section, making the resulting
constraints even weaker.

Let us now briefly summarize the accretion of WIMPs onto a star.
Following~\cite{Press:1985ug,Kouvaris:2007ay,Kouvaris:2010vv}, one can
write the accretion rate as follows
\begin{equation}
F=\frac{8}{3} \pi^2 \frac{\rho_{\text{dm}}}{m} 
\left ( \frac{3}{2 \pi \bar{v}^2} \right )^{3/2} GMR
\bar{v}^2(1-e^{-3E_0/\bar{v}^2})f, 
\label{accretion}
\end{equation}
where $\rho_{\text{dm}}$ is the local dark matter density, $\bar{v}$
is the average WIMP velocity, $M$, and $R$ are the mass and the radius
of the star, $E_0$ is the maximum energy of the WIMP per WIMP mass
that can lead to a capture, and $f$ denotes the probability for at
least one WIMP-proton scattering to take place within the star. In
this expression we have neglected relativistic corrections (which are
very small for regular stars) and possible motion of the star with
respect to the dark matter halo. The latter can reduce slightly the
accretion rate, but not more than an order of magnitude. Since we are
not targeting particular stars like the Sun, we are going to present
results for stars that do not move relatively to the dark matter
halo. However, one can easily get the correct accretion rate in the
case of a moving star by multiplying the accretion rate by
$(\sqrt{\pi}/2)\text{erf}(\eta /\eta)$, where
$\eta=\sqrt{3/2}v_{\odot}/\bar{v}$, with $v_{\odot}$ being the
velocity of the star.

Let us first estimate $E_0$. The recoil energy produced by a
WIMP-proton scattering, is within $0<T<4m_pm/(m+m_p)^2$. Upon assuming
that the distribution over the recoil energies is not very different
from a uniform one, a typical scattering will produce a recoil of
order $\sim 2m_p/m$ (for $m>>m_p$). In order for a WIMP to be captured
(i.e. to become gravitationally bound) after one collision, it must lose
at least the initial kinetic energy it had far out from the star. This
leads to 
\[
E_0 \simeq 2{m_p\over m} {GM\over R}.
\]
This is a conservative estimate because we have
implicitly assumed that the collision will take place at the outskirts
of the star (at the radius $R$), and not somewhere deep inside where the
kinetic energy (and therefore the recoil energy) would be larger. 

For small cross sections, the probability $f$ of at least one
scattering of WIMP in a Sun-like star was estimated in
Ref.~\cite{Press:1985ug} to be
\begin{equation}
f\simeq 0.89{\sigma\over \sigma_{\text{crit}}}\;,
\label{eq:f-def}
\end{equation}
where 
\[
\sigma_{\text{crit}}= {m_pR^2\over M} 
\simeq 4 \times 10^{-36}\text{cm}^2 \left({R \over R_\odot}\right)^2
\left({M\over M_\odot}\right)^{-1},
\]
$M_\odot$ and $R_\odot$ being the mass and the radius of the Sun,
respectively. Note that for another case of interest in what follows,
a supermassive star of a mass $M=15 M_\odot$ and radius
$R=6.75R_\odot$~\cite{Woosley:2002zz}, the critical cross section is
$\sigma_{\rm crit}=1.25 \times 10^{-35}\text{cm}^2$. The probability
$f$ saturates to 1 for $\sigma > \sigma_{\rm crit}$.

We should also mention that in principle, the probability $f$ may
change as a function of time. As the star burns its hydrogen to
helium, WIMPs interacting via spin-dependent cross section passing
through the star meet fewer protons (hydrogen) scatterers to
interact. $He^4$ cannot interact through spin-dependent interactions
with WIMPs. Therefore if we make the simple assumption that the star
has converted all the hydrogen to helium at the end of its hydrogen
stage, the probability $f$ would drop to zero as WIMPs
passing through the star do not find protons to scatter anymore off and be
captured. It is natural to include this effect in the definition of
the critical cross section. In this way one gets for the
spin-dependent case 
\begin{equation}
\sigma_{\text{crit}}
\simeq 5.47 \times 10^{-36}\frac{1}{1-t/t_0}\text{cm}^2 
\left({R \over R_\odot}\right)^2
\left({M\over M_\odot}\right)^{-1},
\label{eq:sigmaCritSD}
\end{equation}
where $t_0$ is the star's lifetime and we have assumed that the initial
composition of the star is $75\%$ hydrogen and $25\%$ helium, i.e. the
one of Big Bang Nucleosynthesis. The critical cross section grows with
time (the probability $f$ decreases).

The WIMP capture rate, Eq.~(\ref{accretion}), can be
written in convenient units as follows,
\begin{equation}
F = 1.1 \times 10^{27} \text{s}^{-1}
\left( \frac{\rho_{dm}}{0.3
  \text{GeV}/\text{cm}^3} \right) \left (\frac{220
  \text{km}/\text{s}}{\bar{v}} \right) \left (\frac{
  \text{TeV}}{m} \right) 
\left({M \over M_\odot}\right) 
\left({R\over R_\odot}\right) 
\left( 1-e^{\frac{-3E_0}{\bar{v}^2}} \right) f , 
\label{accretion2}
\end{equation}
where $f$ is defined by Eqs.~(\ref{eq:f-def}),
and~(\ref{eq:sigmaCritSD}). Making use of Eq.~(\ref{accretion2}) one
can estimate the amount of accreted WIMPs during the lifetime of the
star. A typical Sun-like star will burn hydrogen for 10 billion years
before it becomes a red giant and later a white dwarf. A typical 15
$M_{\bigodot}$ supermassive star burns first hydrogen to helium for
11.1 million years and then helium to carbon, carbon to oxygen etc. in
a much smaller timescale (the last stage before the supernova
explosion is the silicon burning lasting about $20$ days).

\section{Gravitational Collapse of the WIMP-Sphere}

Once the WIMPs are captured by the star, they start to thermalize
through successive collisions with the nuclei inside the star and
after sufficient time are described by the Maxwell-Boltzmann
distribution in the velocity and distance from the center of the
star. The majority of WIMPs then is concentrated within the
radius
\begin{equation} 
r_{th}= \left ( \frac{9T}{8 \pi G \rho_c m} \right )^{1/2} \simeq 2
\times 10^{8} \text{cm}
\left( {m\over {\rm TeV}}\right)^{-1/2} , 
\label{rthermal} 
\end{equation}
where $T$ is the temperature of the star, $\rho_c$ is the core density
and $m$ is the mass of the WIMP. In the last equality we used typical
values for a Sun-like star $T= 1.5 \times 10^7$ K, and
$\rho_c=150$~$\text{g/cm}^3$.  For a supermassive star of mass
$M=15M_{\odot}$, the thermal radius is roughly an order of
magnitude larger (using typical values $T= 3.53 \times 10^7$ K, and
$\rho_c=5.81$~$\text{g/cm}^3$~\cite{Woosley:2002zz}).  

Depending on the mass and the cross section between WIMP and nucleon,
WIMPs might or might not thermalize during the lifetime of the star.
Since our constraints will depend on the WIMP thermalization, we
estimate here the thermalization time. The thermalization of captured
WIMPs can be divided into two stages: at the first stage the WIMPs
oscillate in the star's gravitational potential crossing it twice per
period. This lasts until the WIMP's orbit decreases to the size of the
star. At the second stage, the WIMP moves completely inside the star
on the orbit which shrinks to the thermal radius.

Consider the first stage.  Each time the WIMP crosses the star it has
a chance to collide and lose some energy. The time between collisions
$\Delta t$ is given by half a period of WIMPs oscillation divided by
the ratio of the WIMP cross section to the critical cross section. At
each collision the WIMP loses typically a fraction $2m_p/m$ of its
energy. Averaging over the WIMP trajectory inside the star (assuming
for simplicity that the latter passes through the center) the typical
energy loss is
\[
\Delta E = 2 G M m_p 
\left( {4\over 3 R} - {1\over r}\right),
\]
where $M$ and $R$ are the mass and the radius of the star, and $r$
is the size of the WIMP's orbit. Dividing this energy change by
$\Delta t$ and expressing $r$ in terms of energy gives the
differential equation for the WIMP energy as a function of time,
\[
{d E \over dt} = - {2\sqrt{2} m_p \sigma\over 
\pi G M m^{5/2}} 
\left( {4\over 3} E_* + E\right) |E|^{3/2}, 
\]
where $E_* = GMm/R$ is the binding energy of the WIMP at the
star surface. From this equation the duration of the first stage is 
\begin{equation}
t_1 = {\pi m R^{3/2} \sigma_{\rm crit}\over 
2m_p \sqrt{2GM} \sigma} \int_{\epsilon_0}^1
{d\epsilon \over (4/3-\epsilon)\epsilon^{3/2} }
\sim  {3\pi m R^{3/2} \sigma_{\rm crit}\over 
4m_p \sqrt{2GM} \sigma} \sqrt{{E_*\over |E_0|}},
\label{eq:timet1}
\end{equation}
where $\epsilon_0 = |E_0|/E_*$ is the ratio of the WIMP initial energy
$E_0$ to its binding energy at the star surface. Numerically 
$\epsilon_0 \sim m_p/ m$, and for a solar mass star and
spin-dependent cross section we have 
\begin{equation}
t_1 = 3 \,{\rm yr} 
\left({m\over {\rm TeV}}\right)^{3/2}
\left({\sigma\over 10^{-35}{\rm cm}^2}\right)^{-1}.
\label{eq:t1-final}
\end{equation}
The time becomes longer for larger masses and smaller cross
sections.

At the second stage, the average time between two successive
collisions is $\Delta t=1/(n\sigma v)$, where $n$ is the number
density of the nucleons, $\sigma$ is the WIMP-proton cross section,
and $v=\sqrt{2E/m}$ is the average velocity of the WIMP. Therefore,
the energy as a function of time is determined by the following
equation
\begin{equation}
\frac{dE}{dt}=-n\sigma v \Delta E = 
-2 \sqrt{2}  \rho \sigma \left({E\over m}\right)^{3/2},
\end{equation}
where $\rho$ is the matter density of the star.
Solving this equation gives the time $t_2$ needed to decrease the
energy from $E_{in}$ to $E_f$, 
\begin{equation}
t_2 =\frac{m^{3/2}}{\sqrt{2}\rho\sigma} \left (
\frac{1}{\sqrt{E_f}}-\frac{1}{\sqrt{E_{in}}} \right ).
\label{eq:th_time2}
\end{equation}
Here the initial energy is of order $E_{in}\simeq GM/R$, while the
final energy is determined by the final size of the WIMP cloud.  In
case of cooling to the thermal radius this energy is $E_f=(3/2)kT$,
where $k$ is the Boltzmann constant and $T$ is the temperature in the
core of the star.  One can safely neglect the term depending on
$E_{in}$ in Eq.~(\ref{eq:th_time2}) since $E_{in}\gg E_f$. 

In case of cooling to some fixed radius (e.g. to the radius of a
future white dwarf $\sim 4000$~km, as will be of interest in what
follows) the final energy is determined by this radius and the density
of the star. Assuming typical parameters of a solar mass
star, one gets 
\begin{equation}
t_2 = 0.15 \, {\rm yr} 
\left({m\over {\rm TeV}}\right)
\left({\sigma\over 10^{-35}{\rm cm}^2}\right)^{-1}.
\label{eq:th-time2-sun}
\end{equation}

Once inside either a white dwarf or a neutron star, WIMPs start to
thermalize once again toward the much smaller thermal radius of a
white dwarf (typically a few kilometers for a TeV WIMP) or of a
neutron star (typically a few centimeters for a TeV WIMP). This
process described by the same Eq.~(\ref{eq:th_time2}), but with
different parameters.  If a sufficiently large number of WIMPs has
been accumulated, WIMPs may start self-gravitating and collapse
gravitationally (in the absence obviously of a repulsive force
between them). In this way the formation of a compact star (either a
white dwarf or a neutron star) may trigger the collapse of the WIMP
sphere into a black hole.

The onset of the self-gravitating regime happens when the total mass
of WIMPs inside the thermal radius becomes comparable to the total
mass of the ordinary matter in the same region. This leads to the 
condition 
\[
N \gtrsim \left({T^3\over G^3 m^5 \rho_c}\right)^{1/2} \sim 10^{45} 
\left({m\over {\rm TeV}}\right)^{-5/2},
\]
where we have substituted typical parameters of a white dwarf. For a
neutron star the required number of dark matter particles is 5 orders
of magnitude smaller. 

The self-gravitating WIMP sphere may collapse into a black hole if the
Fermi pressure of the WIMPs cannot counterbalance the gravitational
attraction. The onset of the gravitational collapse occurs when the
potential energy of a WIMP exceeds the Fermi momentum, and therefore
Pauli blocking cannot prevent the collapse anymore. This happens when
\begin{equation}
\frac{GNm^2}{r}>k_F= \left (\frac{3\pi^2N}{V} \right ) = \left (
\frac{9\pi}{4} \right )^{1/3} \frac{N^{1/3}}{r}. 
\end{equation}
In the derivation of the above limit, we have considered that WIMPs
are (semi)-relativistic, which is justified since once WIMPs
self-gravitate themselves, they get closer and closer, building up a
Fermi momentum that eventually corresponds to relativistic
velocities. From the above equation we can deduce the number of WIMPs
needed for the collapse to take place, 
\begin{equation} 
N=\left (\frac{9 \pi}{4}
\right )^{1/2} \left (\frac{m_{\rm Pl}}{m} \right )^3 \simeq 5 \times
10^{48}\left({m\over {\rm TeV}}\right)^{-3},
\end{equation} 
where $m$ is the WIMP mass and $m_{\rm Pl}$ is the Planck mass.

\section{Constraints on Dark Matter}
\label{sec:constr-dark-matt}
Having derived the accretion rate formula for a generic star and the
amount of dark matter needed in order to form a mini-black hole, we
can proceed to the constraints that arise from the requirement that
such mini-black holes are not created inside newly-formed white dwarfs
and neutron stars. We consider two different cases, i.e. constraints
on spin-dependent cross sections from white
dwarfs, and  from neutron
stars.

\subsection{Constraints on Spin-Dependent cross section from White Dwarfs}
\label{sec:spin-dep-case}

Dark Matter WIMPs can have purely spin-independent, or spin-dependent
interactions with nuclei, or even both types at the same time. Due to
the coherence effect, the spin-independent interactions are usually
stronger than the spin-dependent ones. However, there are cases where
the spin-independent cross section is either suppressed or
absent. Such cases arise naturally in models where dark matter
candidates have an axial coupling to gauge bosons.  One characteristic
example is Majorana particles. Majorana fermions scatter off nuclei
without the $N^2$ enhancement mentioned earlier because the amplitudes
of scattering on different nucleons add up incoherently. Since most of
the nucleons within the nuclei come in pairs of opposite spin, the
WIMP interacts effectively only with the unpaired nucleons. However, a
Majorana particle is its own antiparticle, allowing therefore for 
WIMP co-annihilation. The co-annihilation invalidates our constraints
because it destroys the WIMPs before they collapse into a black hole,
unless the annihilation cross section is extremely small. However, in
such a case the spin-dependent elastic cross section should be
parametrically equally small.

The constraints we present in this subsection are valid for dark
matter candidates which predominantly have a spin-dependent cross
section without being Majorana particles. We constrain models of
asymmetric dark matter with WIMPs that have axial couplings to gauge
bosons like $Z$.  Such candidates have been identified and studied
in~\cite{Agrawal:2010fh}. For example, Dirac fermions with predominant
axial coupling to the $Z$ boson have suppressed spin-independent cross
section and dominant spin-dependent one. In asymmetric dark matter
models of this type, although the annihilation cross section is not
suppressed, annihilations are rare due to the asymmetry. In such
cases we can impose constraints competitive to the direct dark matter
search experiments.  

In this subsection we look at potential candidates that have a
spin-dependent cross section with protons which is larger than the
spin-independent cross section. We consider both types of cross
sections compatible with the experimental constraints. In such a case,
WIMPs will be captured by a solar mass star primarily due to their
spin-dependent interactions. When the star turns into a white dwarf,
most of the WIMPs which are inside the white dwarf radius will be
inherited by the white dwarf. 

In order to form a black hole, the accumulated WIMPs have to cool
further. The spin-dependent interactions inside the white dwarf are
suppressed because the latter is composed predominantly of spin-zero
nuclei like He$^4$, C$^{12}$ and O$^{16}$. However, since the white
dwarfs are much denser than the ordinary stars, much smaller
interactions are sufficient for the successful WIMP thermalization. As
we will show below, these interactions may be provided by a small
spin-independent component in the interaction and/or by a small admixture of nuclei
with non-zero spin such as for example C$^{13}$ or O$^{15}$.

Consider as an example the case where there are spin-independent
interactions in addition to the spin-dependent ones. In this case the
dark matter particles are accumulated in a solar mass star mostly due
to their spin-dependent interactions, but once the star collapses to a
white dwarf made of spin-zero nuclei, they thermalize due to
collisions of spin-independent nature. We do not have to
assume that WIMPs interact also with neutrons. It is sufficient to
have spin-dependent interactions with protons only. In fact, the
experimental constraints on spin-dependent cross section of WIMPs with
protons is even less stringent, the strongest bounds coming from 
CDMS~\cite{Ahmed:2008eu}, PICASSO~\cite{Archambault:2009sm}, and
KIMS~\cite{Lee.:2007qn}. The constraint is approximately given by \[
\sigma_{\rm SD,p} < 10^{-36}\text{cm}^2\, 
\left({m\over {\rm TeV}}\right)
\]
at WIMP masses above 1~TeV. Note that stricter constraints on the 
spin-dependent cross section between WIMP-proton from the ICECUBE
collaboration~\cite{Abbasi:2009uz} do not apply in our case since the
constraints are based on annihilation of WIMPs, and we constraint
candidates with spin-dependent cross section but of asymmetric nature,
which effectively have no annihilations because the antiparticles are
simply not present anymore.
 
\begin{figure}[!tbp]
\begin{center}
\includegraphics[width=0.7\linewidth]{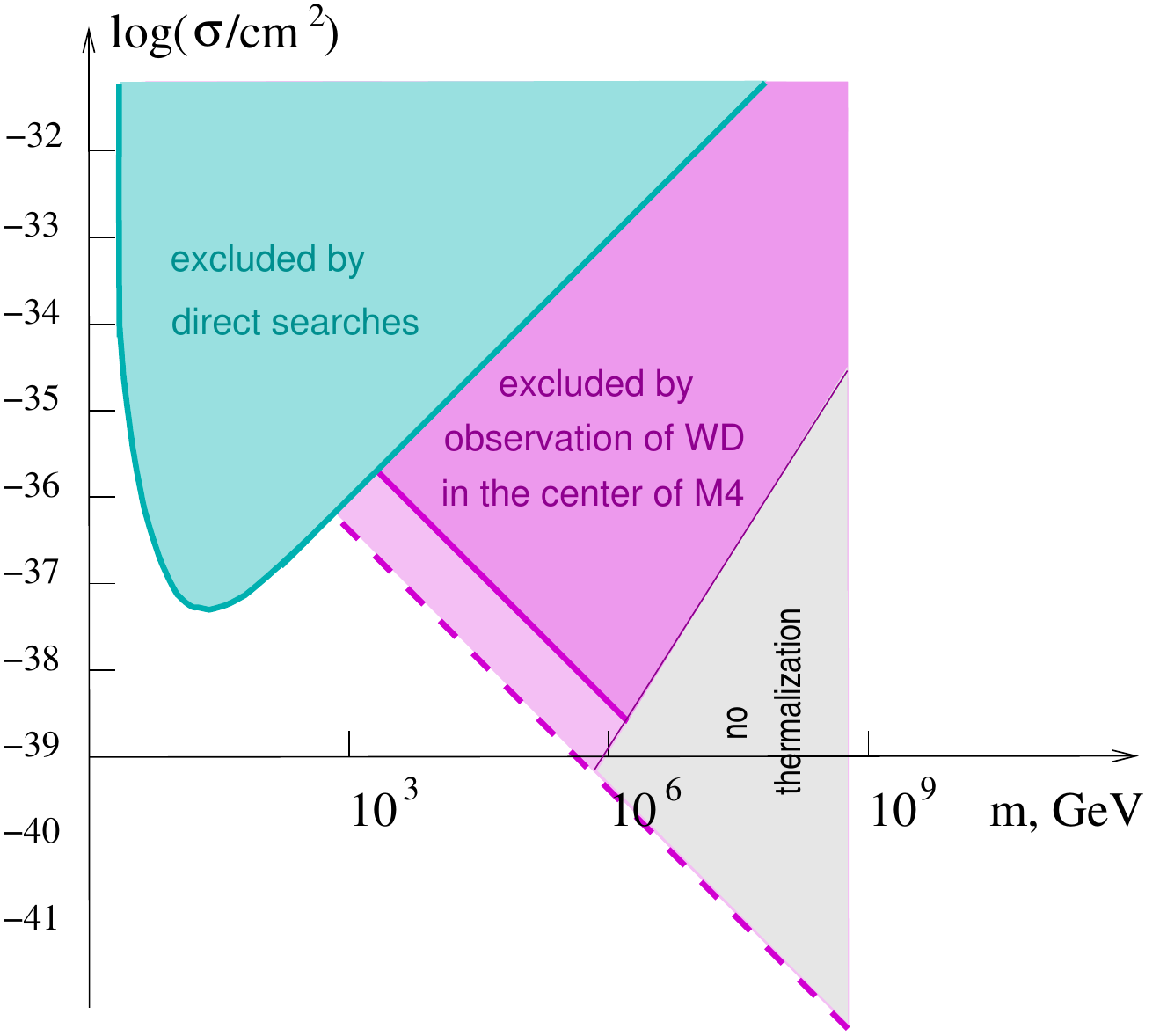}
\caption{Constraints on the spin-dependent WIMP-proton cross
  section. The constraints follow from the existence of old white
  dwarfs in globular clusters with a core dark matter density of
  $10^3$~GeV/cm$^3$ (straight solid line) and $10^4$~GeV/cm$^3$
  (straight dashed line). We also show the constraints from direct
  searches (solid curve).
  }
\end{center}
\end{figure}  
In Fig.~1 we present the resulting constraints on the spin-dependent
cross section, as a function of the WIMP mass. As one can see, these
constraints are competitive with the direct ones at WIMPs masses
of 1 TeV or less, depending on the dark matter density at the
location of the observed white dwarf.  The constraints cannot be
extended to lower masses because no matter what the cross section is,
not enough WIMPs can be accumulated to trigger the collapse.

Let us now estimate the spin-independent cross section required for
thermalization of the WIMPs inside the white dwarf.  Substituting
typical white dwarf parameter in Eq.~(\ref{eq:th_time2}) one finds
\begin{equation}
t_2 = 4 \,{\rm yr} 
\left({m\over {\rm TeV}}\right)^{3/2}
\left({\rho\over 10^8 {\rm g/cm}^3}\right)^{-1}
\left({\sigma\over 10^{-43}{\rm cm}^2}\right)^{-1}
\left({T\over 10^7 K}\right)^{-1/2}.
\label{eq:time-final}
\end{equation}
Given that cold and old white dwarfs (more than several billion years)
have been observed in the inner parts of globular
clusters~\cite{Richer:1997jk}, the thermalization time may safely be
as long as 1~Gyr.  Assuming for simplicity that the white dwarf is
made exclusively of $C^{12}$, this gives \begin{equation}
  \sigma_{SI}^C > 4 \times 10^{-52} \left (\frac{m}{\text{TeV}} \right
  )^{3/2} \left ( \frac{10^7~\text{K}}{T} \right
  )^{1/2}\left({t_0\over {\rm Gyr}}\right)^{-1}~\text{cm}^2,
\label{si_bound2} 
\end{equation} 
where $t_0$ is the maximum time we allow for thermalization (e.g. 1
billion years) and $\sigma_{SI}^C$ is the spin-independent WIMP-carbon
cross section.  The constraint can be rewritten as a constraint for
the WIMP-proton cross section as
\begin{equation}
\sigma_{SI}^p >
8\times 10^{-56}  
\left({m\over {\rm TeV}}\right)^{3/2}
\left({T\over 10^7 K}\right)^{-1/2}
\left({t_0\over {\rm Gyr}}\right)^{-1}~\text{cm}^2,
\label{si_bound1} 
\end{equation}
taking into account that the spin-independent cross section
WIMP-carbon is related to that of WIMP-proton as
$\sigma_{SI}^C/\sigma_{SI}^p \simeq (\mu_C/\mu_H)^26^2$, where $\mu$
are the reduced masses of the respective WIMP-nucleus and $6^2$ is the
coherence enhancement due to the 6 protons of the carbon.

As we have already mentioned, a spin-independent interaction of WIMPs
with nucleons or a small fraction of isotopes of carbon or oxygen with
nonzero spin can play the same role in the thermalization of WIMPs.
By inspection of Eq.~(\ref{eq:time-final}), it is obvious that one can
always trade the cross section for the number density.  This means
that, assuming a TeV WIMP, if thermalization can be achieved for
$\sigma_{SI}^C\sim 10^{-52}\text{cm}^2$, it can equally be achieved by
a spin-dependent cross section $\sigma_{SD}^C=10^{-41}\text{cm}^2$
where $C$ now represents $C^{13}$ with a relative abundance
$C^{12}:C^{13}=1:10^{-11}$.  Several red giants have been seen with a
ratio $C^{12}:C^{13}=4:1$ predicted by the equilibrium processes. This
is because $C^{12}$ from triple-alpha production can meet with
hydrogen in outer shells of the red giant leading to $C^{13}$. In
practice, this means that even with a $10^9$ GeV WIMP an abundance of
$C^{13}$ as low as $C^{12}:C^{13}=100:1$ is sufficient for the
thermalization. Note that the WIMP-$C^{13}$ cross section scales as
$\sim \sigma_{SD}^p \mu{_C}^2/\mu_p^2$ times other nuclear spin
factors, so the abundace of $C^{13}$ can be even $10^4$ smaller
compared to $C^{12}$. We should also emphasize here that $C^{13}$ has
an excess of a neutron, and therefore we have assumed that the WIMP
couples equally to protons and neutrons. If the WIMP couples only to
protons but not neutrons, another isotope has to be considered, such
as $O^{15}$ or $N^{13}$.

For our constraints to be valid, we have to make sure that the black
hole formed inside the white dwarf can eat/destroy the star
within at most 1 billion years (smaller than the age of the older
white dwarfs observed in globular clusters).  First note that once the
WIMPs form a black hole inside a white dwarf, regular nuclear matter
starts falling in it. This process has been considered in
Ref.~\cite{Giddings:2008gr}. It has been argued that for black holes
of a size exceeding the atomic size the accretion proceeds in the
Bondi regime which is characterized by a quasi-stationary matter flow
into the black hole. The total rate of matter accretion can be
expressed in terms of the matter density and sound speed far from the
black hole. The accretion rate is proportional to the square of the
black hole mass, so the change of the black hole mass with time is
described by the equation
\[
M(t) = {M_0\over 1-t/t_*},
\]
where $M_0$ is the initial black hole mass and $t_*$ is the
characteristic time scale over which the star is destroyed, 
\[
t_* = {c_s^3\over \pi G^2 \rho_c M_0}.
\]
Here $c_s$ and $\rho_c$ are the sound speed and the density 
in the core of the white dwarf, respectively. 
Numerically, the time $t_*$ is
\[
t_* \sim  8\cdot 10^3 {\rm yr} 
\left( {M_0\over 10^{-12}M_\odot}\right)^{-1},
\]
where we have estimated the value $c_s=0.03 c$ which is consistent
with~\cite{Giddings:2008gr} and the core density $\rho =
10^8$~g/cm$^3$. The heavier is the WIMP, the smaller is the amount of
dark matter accumulated, and the smaller is the mass of the black hole
created when the WIMPs collapse. For black holes made of collapsing
WIMPs up to masses $10^9$ GeV or slightly lower, the time it takes
for the full destruction of the star (i.e. the time it takes for the
black hole to eat the whole star) is less than a billion years.

The grey ``no thermalization" area in Fig.~1, (see
Eq.~(\ref{eq:t1-final}) and the discussion following) corresponds to
cases where the captured WIMPs in the progenitor did not have enough
time to settle within the radius of a white dwarf, meaning that
although most of them will be gravitationally bound to the white dwarf
when it is formed, their orbits do not necessarily intersect with it,
and therefore there is the danger of not thermalizing and collapse
gravitationally.  Although in principle in such a case WIMPs might not
collapse gravitationally, the situation is far from clear and we leave
this for a future study.  This is because WIMP-WIMP interactions can
redistribute the angular momentum of the WIMPs and might eventually
lead a large fraction of them to be captured fast by the white
dwarf. In addition if the white dwarf is part of a binary system,
there is again the possibility of allowing WIMPs to intersect with the
white dwarfs.  We should emphasize that to the left of the grey area,
there is no ambiguity about the outcome, since WIMPs have enough time
to concentrate in a radius smaller than that of the would-be white
dwarf, and therefore once the white dwarf is formed, they are already
inside.

\subsection{Constraints on Spin-Dependent cross section from Neutron Stars }
\label{sec:spin-dependent-case}

Constraints on spin-dependent cross section can be imposed also
directly by accretion of WIMPs into neutron
stars~\cite{Goldman:1989nd}, or by accretion of WIMPs in a progenitor
supermassive star that later collapses to a neutron star.  We consider
supermassive stars that accrete dark matter during the hydrogen
burning stage. At other stages of the star evolution the accretion is
negligible since the star is mainly composed of spin-zero nuclei
like $He^4$, $C^{12}$, $O^{16}$, etc., and also these stages are much
shorter.  After the last (silicon) stage, the supermassive star
collapses to a neutron star. The supernova explosion cannot blow out
the accumulated WIMPs because the spin-dependent interaction with
silicon is zero. After the formation of the neutron star, the
accumulated WIMPs are located mostly outside the neutron star, simply
because the thermal radius of a supermassive star where WIMPs
concentrate during the hydrogen-burning stage is several orders of
magnitude larger than the radius of a neutron star. Upon assuming that
WIMP-WIMP interactions or a binary system can redistribute the angular
momentum and randomize the velocity of the WIMPs trapped by the
supermassive progenitor, after the neutron star's formation, the WIMPs
are captured by it very quickly~\cite{Kouvaris:2010vv}. Inside the
neutron star the WIMPs continue to scatter on neutrons until coming to
a thermal equilibrium with the star. We should emphasize that we
consider asymmetric WIMPs with spin-dependent cross section with both
protons and neutrons here.

There is an issue to address in the first place which is why to
consider constraints which are due to the WIMP accumulation in the
progenitor of a neutron star and not directly those which arise due to
capture by the neutron star itself. The latter constraints have been
first considered in~\cite{Goldman:1989nd}.  For the spin-independent
cross sections compatible with the experimental limits, a neutron star
can accumulate a similar amount of WIMPs as a supermassive star. This
is because the spin-independent cross sections consistent with the
experiments are many orders of magnitude smaller than the critical
cross section of a massive star, but not necessarily smaller compared
to the critical cross section of a neutron star. Given the fact that
massive stars live (and accrete dark matter) for a much smaller time
compared to neutron stars, the latter are more efficient in
accumulating WIMPs overall.  However, in the case of spin-dependent
interaction, experimental limits on the cross section are comparable
to the critical cross section of a supermassive star. In such a case,
a massive star can accumulate more WIMPs in the 10 million years of the
hydrogen-burning stage than a neutron star would capture in 10 billion
years.  Therefore, the constraints resulting from the former process
are stronger.

\begin{figure}[!tbp]
\begin{center}
\includegraphics[width=0.7\linewidth]{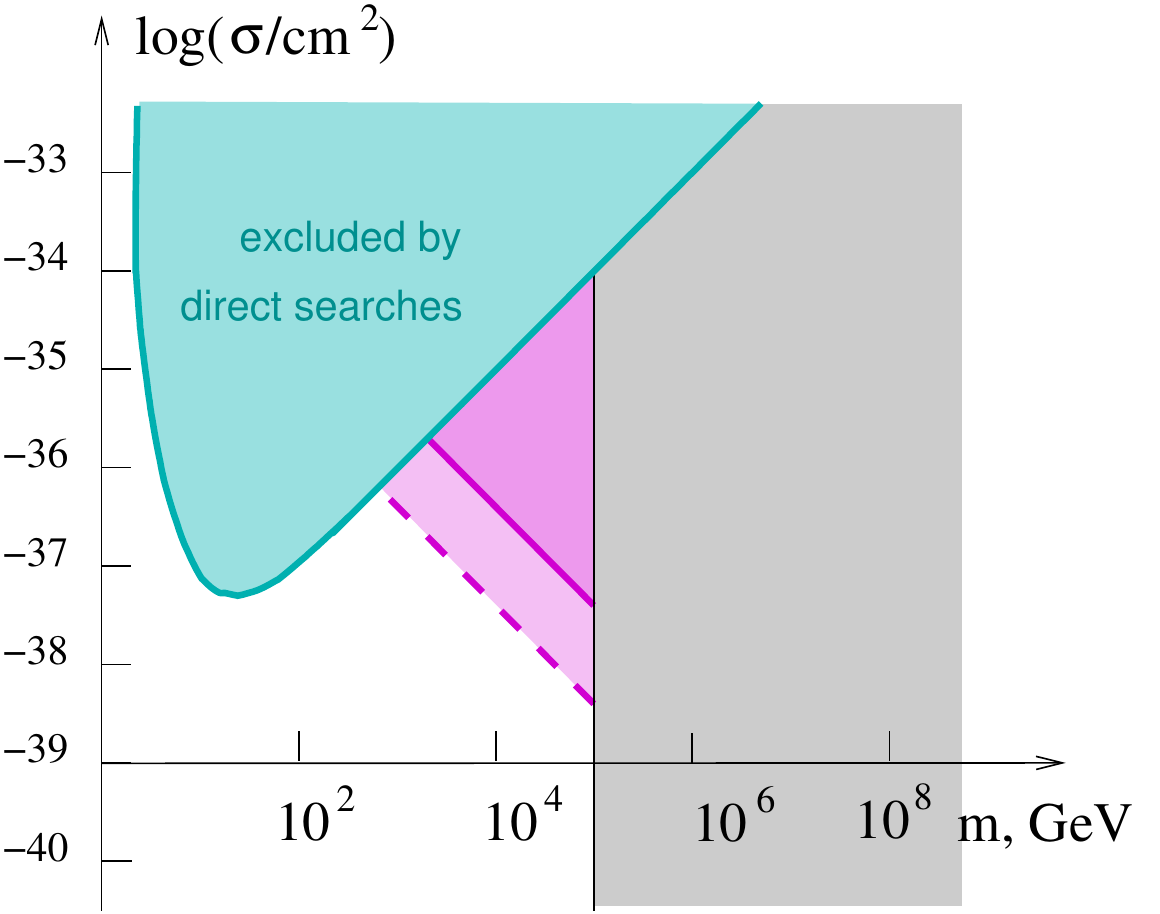}
\caption{Constraints on the spin-dependent WIMP-nucleon cross section
  that follow from the existence of neutron stars in globular
  clusters. The grey area is excluded by direct accretion of WIMPs
  (for 10 Gyr) in an old neutron star with local dark matter density
  $10^3$ GeV$/\text{cm}^3$. The purple area shows potential exclusion
  upon making an assumption regarding the distribution of WIMP
  velocities (see text), with the solid (dashed) line corresponding to
  local dark mater density $10^3~(10^4)$ GeV$/\text{cm}^3$. We also
  show the exclusion area by direct dark matter search experiments.}
\end{center}
\end{figure}
Several pulsars have been detected in globular clusters close to (or
inside) the core~\cite{Camilo:2005aa}. In addition neutron stars older
than 10 billion years have also been observed (as for example the
B1620-26 in M4). In Fig.~2 we present the constraints on the
spin-dependent WIMP-nucleon cross section that arise from the
non-gravitational collapse of WIMPs inside neutron stars in
dark-matter-rich environments such as the globular cluster M4. The
grey rectangular area corresponds to exclusion due to direct neutron
star accretion of WIMPs for roughly 10 billion years (with
$\rho_{dm}=10^3$ GeV$/\text{cm}^3$) similarly to what previously has
been derived
in~\cite{Goldman:1989nd,Bertone:2007ae,Kouvaris:2010vv}. The area
appears as a rectangular due to the fact that it excludes cross
sections down to the critical cross section of a neutron star ($\sim
10^{-45}~\text{cm}^2$~\cite{Kouvaris:2010vv}). The triangle area in
purple is excluded (if one assumes the redistribution of WIMP
velocities we have already discussed) due to WIMPs that have been
accreted by the progenitor during its lifetime, and collapse in the
core of the neutron star once they are ``sucked'' inside the neutron
star.  As one can see from the figure, the existence of a neutron star
in the core of a globular cluster can impose in principle constraints
competitive to those from the Earth based dark matter search
experiments for WIMP masses above 3 to 10 TeV depending on the local
dark matter density.  Our constraints become better than the direct
ones when the mass increases further, since at larger WIMP masses
fewer particles are needed for the collapse, whereas for the Earth
based experiments it simply means fewer events in the detector.

In order to make sure that our constraints are valid, we have to
estimate the time it takes for the WIMPs to collapse once the neutron
star is formed. As we mentioned earlier, the time it takes for the
WIMPs accumulated by the progenitor to be captured by the neutron star
is negligible. Once inside the neutron star, WIMPs have to lose energy
via collisions in order to concentrate in the center and
collapse. This time has been estimated in~\cite{Kouvaris:2010vv} and
for all relevant cases is at most of the order of a year. In addition,
we have checked that for the whole relevant range of WIMP masses,
self-gravitation of the WIMPs sets on even before they all concentrate
within the thermal radius of a neutron star.  This means that the
collapse is even faster.

We should emphasize that if a mechanism for randomizing the velocities
of the WIMPs is operating (and therefore our constraints are valid), we
exclude parameter space (purple triangle) that is not excluded by the
Earth based experiments or by the gravitational collapse of WIMPs
accumulated by the neutron star itself.  Such a constraint can also be
to some extent more robust than the one derived previously because no
assumption is made regarding the age of the neutron star (as long as
there is enough time for the black hole to destroy the star). On the
contrary, in the constraints that come from direct accretion of WIMPs
in the neutron star, one has to know with enough precision the age of
the star in order to estimate the amount of WIMPs accreted.

\section{Conclusions}

We derived constraints on the spin-dependent cross section of
asymmetric fermionic dark matter WIMPs based on the existence of white
dwarfs and neutron stars in globular clusters. Our constraints are
competitive to direct dark matter search experiments, excluding a large
parameter space of cross sections and masses as low as TeV (or
slightly lower than TeV). 

In the case of white dwarfs, we were able to exclude a range of
spin-dependent cross sections and WIMP masses because for these
parameters, WIMPs that have been captured during the lifetime of the
progenitor have enough time to concentrate within the core of the star
that is inherited by the white dwarf, and eventually collapse
gravitationally forming a black hole that destroys the star. This
constraint is robust in the sense that it depends only on the local
dark matter density of the globular cluster and no other
hypothesis. We demonstrated that asymmetric WIMP candidates with only
spin-dependent interactions with masses even lower than one TeV,
trapped during the lifetime of the progenitor can easily thermalize
inside the white dwarf due to expected small abundances of isotopes of
carbon or oxygen that carry spin.

If we now relax the strict condition of thermalization, i.e. to assume
that WIMPs are gravitationally trapped by the progenitor but are not
necessarily confined within the radius of a white dwarf, our
constraints can be extended to higher masses and lower cross
sections. However, in this case we have to make an extra assumption of
a mechanism that redistributes the WIMP velocities in order for WIMPs
which are on orbits around the white dwarf to intersect with it. Such
a mechanism can be possibly provided by binaries or WIMP-WIMP
interactions.  Further investigation of this possibility is needed.

In the case of neutron stars, we exclude an area of WIMP masses and
cross sections due to direct accretion of WIMPs in the neutron
star. The constraints depend only on the local dark matter density of
the globular cluster and the age of the neutron star. Upon assuming an
extra mechanism of WIMP velocity redistribution, extra parameter space
may be excluded down to the TeV scale.

\end{document}